\documentclass[prd,twocolumn,showpacs,preprintnumbers,superscriptaddress,floatfix,nofootinbib]{revtex4-1}
\usepackage{graphicx,,booktabs}
\usepackage{amssymb}
\usepackage{amsmath,txfonts}
\usepackage{graphicx}
\usepackage{dcolumn}
\usepackage{color}
\usepackage{bm}
\usepackage[colorlinks,
            citecolor=blue,
            anchorcolor=green,
            menucolor=orange,
            linkcolor=red,
            filecolor=red,
            runcolor=pink,
            urlcolor=blue,
            frenchlinks=red]{hyperref}
\usepackage{multirow}

\begin{document}

\title{\boldmath Production of
the $\eta_{1}(1855)$ through
kaon induced reactions under the assumptions that it is a molecular or a hybrid state}
\author{Xiao-Yun Wang}
\email{xywang@lut.edu.cn}
\affiliation{Department of physics, Lanzhou University of Technology,
Lanzhou 730050, China}
\affiliation{Lanzhou Center for Theoretical Physics, Key Laboratory of Theoretical Physics of Gansu Province, Lanzhou University, Lanzhou, Gansu 730000, China}
\author{Fan-Cong Zeng}
\affiliation{University of Chinese Academy of Sciences, Beijing 100049, China}
\author{Xiang Liu}
\email{xiangliu@lzu.edu.cn}
\affiliation{School of Physical Science and Technology, Lanzhou University, Lanzhou 730000, China}
\affiliation{Research Center for Hadron and CSR Physics, Lanzhou University and Institute of Modern Physics of CAS, Lanzhou 730000, China}
\affiliation{Lanzhou Center for Theoretical Physics, Key Laboratory of Theoretical Physics of Gansu Province, Lanzhou University, Lanzhou, Gansu 730000, China}
\affiliation{Frontiers Science Center for Rare Isotopes, Lanzhou University, Lanzhou 730000, China}
\affiliation{Joint Research Center for Physics, Lanzhou University and Qinghai Normal University, Xining 810000, China}
\begin{abstract}
By the reaction of kaon interacting with a proton, we investigate the production of the newly observed $\eta _{1}(1855)$ predicted in the picture of the $K\bar{K}_1(1400)$ molecular state and hybrid state. The total and differential
cross sections of the concrete $K^{-}p\to\eta _{1}(1855)\Lambda$ reaction are calculated. Taking the partial decay width of the $\eta _{1}$ to $K\bar{K}^{\ast }$ as 0.9 MeV and 98.1 MeV, the minimum cross section of the  $\eta _{1}(1855)$ production via the $K^{-}p$ reaction can reach up 0.59 $nb$ and 63.8 $nb$ at the center of mass energies $W\simeq 3.5$ GeV, respectively. The differential cross sections for the $\eta _{1}(1855)$ production at the different center of mass energies are also available. Furthermore, we present the Dalitz processes of $2\rightarrow 3$ and $2\rightarrow 4$, and initially discuss
the feasibility of finding out the $\eta _{1}(1855)$ in experiments like J-PARC.
\end{abstract}

\maketitle

\section{Introduction}

Until now, the BESIII Collaboration has collected the world largest data sample
of $J/\psi$, which provide a good platform to carry out the study of the light hadron spectrum \cite{BESIII:2020nme}.
Recently, by analyzing the partial wave of the $J/\psi \rightarrow
\gamma \eta \eta ^{\prime }$ decay, the BESIII Collaboration \cite%
{BESIII:2022qzu,BESIII:2022riz} reported the observation of the $\eta _{1}(1855)$ in the $\eta \eta ^{\prime }$ invariant mass spectrum with a
significance of  19$\sigma$, which has
quantum
number $I^{G}(J^{PC})=0^{+}(1^{-+})$. And its mass and width are
\begin{eqnarray}
m=1855\pm 9_{-1}^{+6}\text{ MeV}, \quad
\Gamma =188\pm 18_{-1}^{+6}\text{ MeV,}
\end{eqnarray}
respectively. Obviously, the observed $\eta _{1}(1855)$ cannot be grouped into the conventional hadron, which is a good candidate of exotic states as indicated by some theoretical groups, where
the $\eta _{1}(1855)$ was interpreted as a hybrid state \cite%
{Chen:2022qpd,Qiu:2022ktc,Shastry:2022mhk}, a molecular state \cite%
{Dong:2022cuw,Yang:2022lwq}, and a tetraquark state \cite{Wan:2022xkx}.

At present, the $\eta _{1}(1855)$ was only observed in the radiative decay $J/\psi \rightarrow
\gamma \eta \eta ^{\prime }$, confirming this observation by other processes is an interesting experimental topic, which will be helpful to establish this exotic state in experiment and may provide some valuable information to reflect its inner structure. Under different assignments to the $\eta _{1}(1855)$ \cite{Chen:2022qpd,Dong:2022cuw}, the strong decays of the
$\eta _{1}(1855)$ were studied, where the $\eta _{1}(1855)$ has strong interaction with the $K^{(*)}\bar{K}^{(*)}$ channels. This result inspires us to propose that the reaction of kaon and proton can be applied to study the production of the
$\eta _{1}(1855)$.

By checking the Particle Data Group \cite{pdg2021}, we may find that the $\bar{K}p$ reaction is an ideal process to explore the light hadron spectrum. Taking the $\phi(1020)$ as an  example, the $\phi(1020)$ was firstly found in the $K^-p\to \Lambda K\bar{K}$ reaction \cite{Schlein:1963zz}. When investigating the production problem of the
$\eta _{1}(1855)$, we naturally pay attention to the $\bar{K}p$ reaction, which is an effective approach among these possible production processes. There are available experiments at OKA@U-70~\cite{Obraztsov:2016lhp} and SPS@CERN~\cite%
{Velghe:2016jjw}, and J-PARC~\cite{Nagae:2008zz,jparc}.
For answering whether the $\eta_1(1855)$ can be accessible at  these facilities, we should have a realistic study of the production of the $\eta_1(1855)$ via the $K^{-}p$ reaction.

Table \ref{width} shows the partial decay width of $\eta_{1}$ predicted based on the hybrid state \cite{Chen:2022qpd} and molecular state \cite{Dong:2022cuw} assumptions, respectively. We note that in the molecular state model hypothesis \cite{Dong:2022cuw}, the width of the  $\eta_1(1855)$ decays to $K\bar{K}^{\ast }$ is only 0.9 MeV, while in the hybrid state model \cite{Chen:2022qpd} the width of the  $\eta_1(1855)$ decays to $K\bar{K}^{\ast }$ reaches up to 98.1 MeV. Such a large decay width difference may lead to the difficulty of finding the $\eta_1(1855)$ through $\bar{K}p$ scattering experiments, so it is interesting to study the $K^-p\to \eta_1(1855)\Lambda$ scattering process, which will help us understand and clarify the nature of the $\eta_1(1855)$. In this work, we calculate the production of $K^-p\to \eta_1(1855)\Lambda$ with a
$t$-channel $K/K^{\ast }$ exchange, where the effective Lagrangian approach and Regge model are adopted. In the following section, the details will be given.

\begin{table}[h]
\renewcommand\tabcolsep{0.01cm}
\renewcommand{\arraystretch}{1.50}
\caption{ The calculated partial decay widths (in units of MeV) of the $\protect\eta _{1}(1855)$ under the hybrid \cite{Chen:2022qpd} and molecular \cite{Dong:2022cuw} pictures. }
\label{width}{\footnotesize \centering
\begin{tabular}{c|cc cccc}
\toprule[1.0pt]
\toprule[1.0pt]
\multirow{2}*{\ ~~~channel~~~} & \multicolumn{2}{c }{~$\eta _{1}(1855)$ $(J^{PC}=1^{-+})$}    \\
\cline{2-3}
& ~~~ Hybrid state \cite{Chen:2022qpd}~~~ & ~~~Molecule state \cite{Dong:2022cuw}~~~   \\ \hline
~ $K^* \bar{K}^* $ ~ &--- & 26.3   \\ \hline
~ $K  \bar{K}  $ ~ &--- & 0    \\ \hline
~ $K  \bar{K}^*  $ ~ & $98.1$  & 0.9  \\ \hline
~ $a_1 \pi  $ ~ &--- & 9.2   \\ \hline
~ $f_1 \eta  $ ~ &--- & 0.2    \\ \hline
~ $ \eta  \eta^{\prime }  $ ~ &$0.7\sim1.8$ & 26.9    \\ \hline
~ $ \sigma  \omega  $ ~ &--- & 0    \\ \hline
~ $ \rho \rho  $ ~ &--- & 0.04   \\ \hline
~ $ \pi \rho  $ ~ &--- & 0     \\ \hline
~ $ \eta\omega $ ~ &--- & 0    \\ \hline
~ $  \omega  \omega $ ~ &--- & 0.01    \\ \hline
~ $  \omega  \phi $ ~ &--- & 0.4    \\ \hline
~ $  K_{1}(1270)  \bar{K} $ ~ &30.4 & ---    \\ \hline
~ $ K  \bar{K}^*\pi $ ~ &--- & 105.0    \\
\bottomrule[1.0pt]
\bottomrule[1.0pt]
\end{tabular}
}
\end{table}

This paper is organized as follows. After the introduction, we present the formalism
including Lagrangians and amplitudes of the $\eta_1(1855)$ production in Sec. \ref{sec2}. The numerical results are discussed in Sec. \ref{sec3}, followed by a brief summary in Sec. \ref{sec4}.

\section{The production of the $\eta_{1}(1855)$ via the $\bar{K}p$ reaction}\label{sec2}

\begin{figure}[b]
\begin{center}
\includegraphics[scale=0.65]{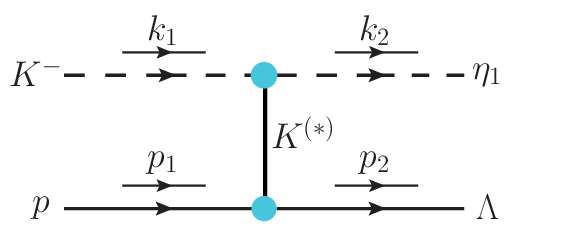}
\end{center}
\caption{Schematic diagram for the $t$-channel contribution to the $K^{-}p\rightarrow \protect%
\eta _{1}(1855)\Lambda $ reaction.}
\label{Fig: Feynman}
\end{figure}

The exotic states $\eta_{1}(1855)$ $(\equiv \eta _{1})$ can be produced via the $\bar{K}p$ reaction, where the diagram is shown in Fig.~\ref{Fig: Feynman} when the $t$-channel exchange of the $K/K^{\ast }$ meson is considered.
In the present work, the contributions from the $u$- and $s$-channel are negligibly small because they are strongly suppressed due to the reason that the two channels proceed with the annihilation of a quark pair in the initial state and the creation of an additional light quark pair in the final state. Moreover, since the total cross section may be reduced after introducing Reggeized treatment for baryon exchange \cite{Wang:2017qcw}, the contributions from the $s$- and $u$-channel with baryon exchange will not be included in this work.

For kaon induced production of the $\eta _{1}(1855)$, the relevant Lagrangians for the $t$-channel
read as below \cite{Wan:2015gsl,Ryu:2012tw,Liu:2008qx,Xiang:2020phx}%
\begin{eqnarray}
\mathcal{L}_{\eta _{1}KK} &=&-ig_{\eta _{1}KK}[(\partial ^{\mu }K)\bar{K}-(\partial ^{\mu }\bar{K})K]\eta _{1\mu
}, \\
\mathcal{L}_{KN\Lambda } &=&ig_{KN\Lambda }\bar{N}\gamma _{5}\Lambda K{\ +}%
\text{ H.c.}, \\
\mathcal{L}_{\eta _{1}K^{\ast }K} &=&\frac{g_{\eta _{1}K^{\ast }K}}{m_{\eta _{1}}}\epsilon _{\alpha \beta \mu \nu
}\partial ^{\beta }\eta _{1}^{\alpha }\partial ^{\nu }K^{\ast \mu
}K, \\
\mathcal{L}_{K^{\ast }N\Lambda } &=&-g_{K^{\ast }N\Lambda }\bar{N}\left( %
\rlap{$\slash$}K^{\ast }-\frac{\kappa _{K^{\ast }N\Lambda }}{2m_{N}}\sigma
_{\mu \nu }\partial ^{\nu }K^{\ast \mu }\right) \Lambda +\text{H.c.}~,
\end{eqnarray}%
where $\epsilon _{\alpha \beta \mu \nu }$ is the Levi-Civita tensor. $\eta
_{1}$, ${K}$, $K^{\ast }$, $N$ and $\Lambda $ stand
for the fields of the $\eta _{1}(1855)$, $K$, $K^{\ast }$
meson, nucleon and $\Lambda$, respectively. The coupling constant $%
g_{KN\Lambda }=-13.24$ can be determined \cite{Wan:2015gsl} by the
SU(3) flavor symmetry relation \cite{Oh:2006hm,Oh:2006in}. Moreover, in the
Nijmegen potential~\cite{Stoks:1999bz}, the calculated results show that
the values of the coupling constants $g_{K^{\ast }N\Lambda }$ and $\kappa
_{K^{\ast }N\Lambda }$ are -4.26 and 2.66, respectively. From Table~\ref{width}, it is noted that based on the molecular state assumption, the decay widths of the  $\eta _{1}(1855)$ to $K\bar{K}$ and $K\bar{K}^{\ast }$ are 0 and 0.9 MeV, respectively, resulting in the  coupling constants $g_{\eta _{1}KK}\simeq 0$ and $g_{\eta _{1}K^{\ast }K}\simeq 0.73 $. In addition, by assuming that $\eta _{1}$ is a hybrid state \cite{Chen:2022qpd}, the partial decay width of the $\eta _{1}(1855)$ to $K\bar{K}^{\ast }$ given in the Ref. \cite{Chen:2022qpd} is 98.1 MeV, and the coupling constant $g_{\eta _{1}K^{\ast }K}\simeq 7.6$ is determined by this partial width,
which is about an order of magnitude higher than the coupling constant obtained by using the partial width in the Ref. \cite{Dong:2022cuw}.
For the width of the  $\eta _{1}(1855)$ decay to $K\bar{K}$, no corresponding value is given in the Ref. \cite{Chen:2022qpd}. Therefore, our subsequent calculations will only consider the cross section of the  $\eta _{1}(1855)$ through $t$-channel $K^{\ast }$ exchange. Due to the large difference in the width of the $\eta _{1}(1855)$ decaying to $K\bar{K}^{\ast }$ given by different models, this will lead to differences in the final cross section size, which will ultimately affect the difficulty of detecting the $\eta _{1}(1855)$ via the $K^{-}p$ reaction.

With the above Lagrangians, the amplitude of the $\eta _{1}(1855)$ production via $t$-channel $K^{\ast }$ exchange
in the $K^{-}p$ scatterings can be written as%
\begin{eqnarray}
i\mathcal{M}_{K^{\ast }} &=&i\frac{g_{\eta_{1}K^{\ast }K}}{%
m_{\eta_{1}}}g_{K^{\ast }N\Lambda }F(q^{2})\bar{u}_N(p_{2})
\notag \left( \gamma _{\nu }-\frac{\kappa _{K^{\ast }N\Lambda }}{2m_{N}}%
\gamma _{\nu }\rlap{$\slash$}q_{K^{\ast }}\right)  \notag 
\\
&&\times
\epsilon_{\alpha\beta\mu\xi}\frac{\mathcal{P}^{\mu \nu }}{%
t-m_{K^{\ast }}^{2}}k_{2}^{\beta} \epsilon _{\eta_{1}}^{\alpha
}(k_{1}-k_{2})^{\xi}  \label{AmpT2}
\end{eqnarray}%
with%
\begin{equation}
\mathcal{P}^{\mu \nu }=-i\left( g^{\mu \nu }+q_{K^{\ast }}^{\mu }q_{K^{\ast
}}^{\nu }/m_{K^{\ast }}^{2}\right) ,
\end{equation}%
where $\epsilon _{\eta_{1}}$ is the polarization vector of the $\eta _{1}(1855)$, and $\bar{u}_N$ or $u_\Lambda $ is
the Dirac spinor of nucleon or $\Lambda $ baryon. For the $t$ channel meson
exchange \cite{Liu:2008qx}, the form factor $F(q^{2})=(\Lambda
_{t}^{2}-m^{2})/(\Lambda _{t}^{2}-q^{2})$ is adopted. Here, $%
t=q^{2}=(k_{1}-k_{2})^{2}$ is the Mandelstam variables. The cutoff $\Lambda
_{t}$ in the form factor is the only free parameter. In
Sec. \ref{sec3}, we discuss how to fix it when presenting the results.


Usually, the Regge trajectory{\ model is successfully applied to analyze the hadron
production at high energy \cite%
{Wan:2015gsl,Wang:2015xwa,Haberzettl:2015exa,Wang:2015hfm,Ozaki:2010wp,Wang:2017qcw,Wang:2017plf}%
. The Reggeization can be done by replacing the $t$-channel propagator in
the Feynman amplitudes~(see Eqs. (\ref{AmpT2})) }with the Regge
propagator
\begin{eqnarray}
\frac{1}{t-m_{K^{\ast }}^{2}} &\rightarrow &\left(\frac{s}{s_{scale}}%
\right)^{\alpha _{K^{\ast }}(t)-1}\frac{\pi \alpha _{K^{\ast }}^{\prime }}{%
\Gamma \lbrack \alpha _{K^{\ast }}(t)]\sin [\pi \alpha _{K^{\ast }}(t)]}.
\end{eqnarray}%
The scale factor $s_{scale}$ is fixed at 1 GeV. In addition, the Regge
trajectories $\alpha _{K^{\ast }}(t)$ read as \cite%
{Ozaki:2010wp}%
\begin{equation}
\alpha _{K^{\ast
}}(t)=1+0.85(t-m_{K^{\ast }}^{2}).\quad \ \
\end{equation}%
It is necessary to note that no additional parameter is introduced after applying the Reggeized
treatment.

\section{Numerical results}\label{sec3}

\subsection{Cross section}

With the preparation shown in the previous section, the cross section of the $%
K^{-}p\rightarrow \eta_{1}(1855)\Lambda $ reaction can be calculated. The differential
cross section in the center of mass (c.m.) frame is written as
\begin{equation}
\frac{d\sigma }{d\cos \theta }=\frac{1}{32\pi s}\frac{\left\vert \vec{k}%
_{2}^{{~\mathrm{c.m.}}}\right\vert }{\left\vert \vec{k}_{1}^{{~\mathrm{c.m.}}%
}\right\vert }\left( \frac{1}{2}\sum\limits_{\lambda }\left\vert \mathcal{M}%
\right\vert ^{2}\right) ,
\end{equation}%
where $s=(k_{1}+p_{1})^{2}$ is defined, and $\theta $ denotes the angle of the outgoing
$\eta _{1}$ meson relative to the $K$ beam direction in the
c.m. frame. $\vec{k}_{1}^{{~\mathrm{c.m.}}}$ and $\vec{k}_{2}^{{~\mathrm{c.m.%
}}}$ are the three-momenta of initial $K$ beam and final $\eta _{1}$, respectively.

\begin{figure}[h]
\begin{center}
\includegraphics[scale=0.4]{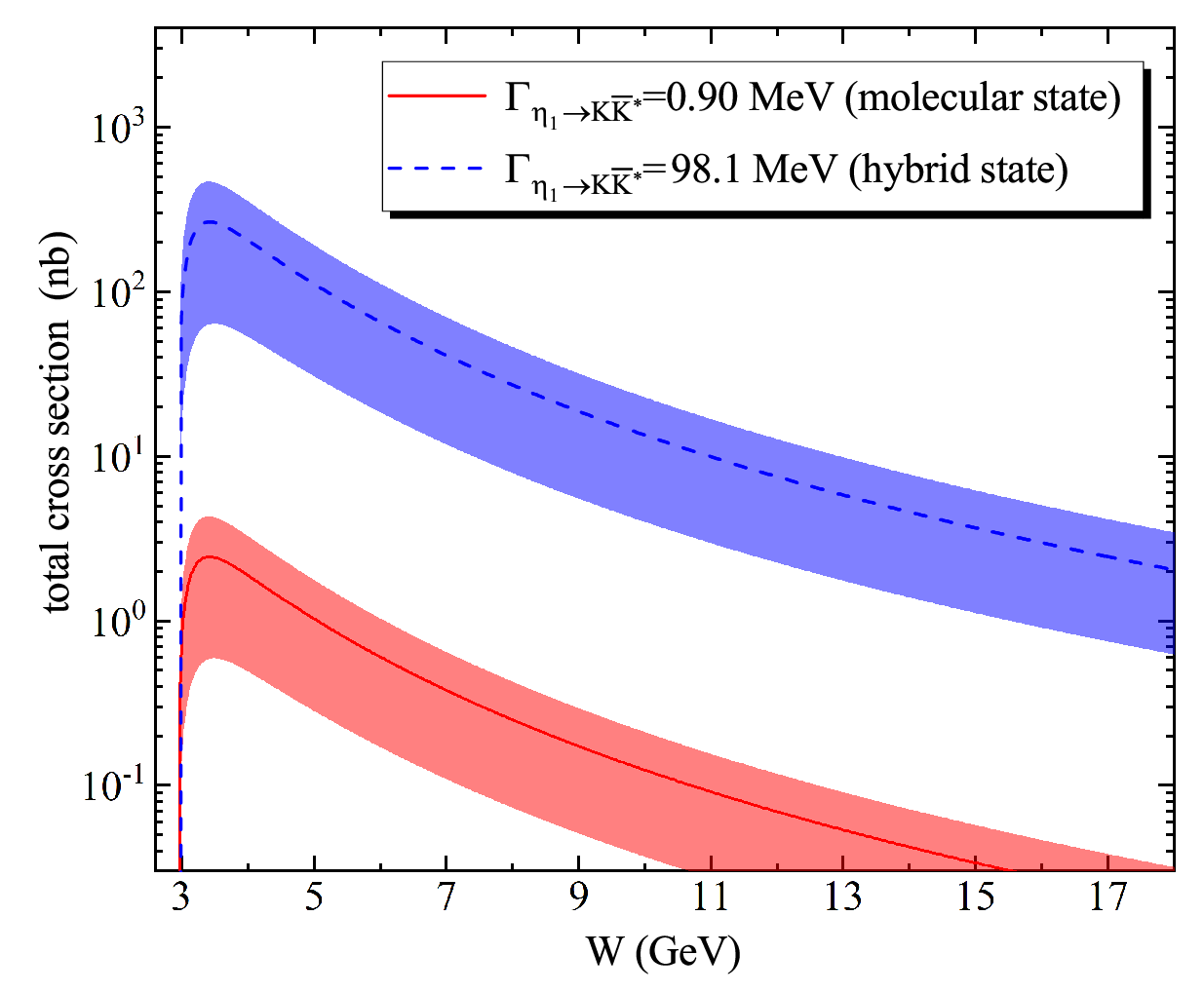}
\end{center}
\caption{(color online).  The energy dependence of the total cross section for
production of the $\protect\eta _{1}(1855)$ through $t$ channel with cutoff $\Lambda _{t}=1.6\pm 0.3$ GeV. The
Full (red) and dashed (blue) lines are the corresponding results when the partial width of $\eta _{1}$ decay into $KK^{\ast}$ is taken as 0.9 MeV and 98.1 MeV, respectively. The bands stand for the error bar of the cutoff $%
\Lambda _{t}.$}
\label{Fig:total}
\end{figure}

\begin{figure}[htbp]
\centering
\includegraphics[scale=0.41]{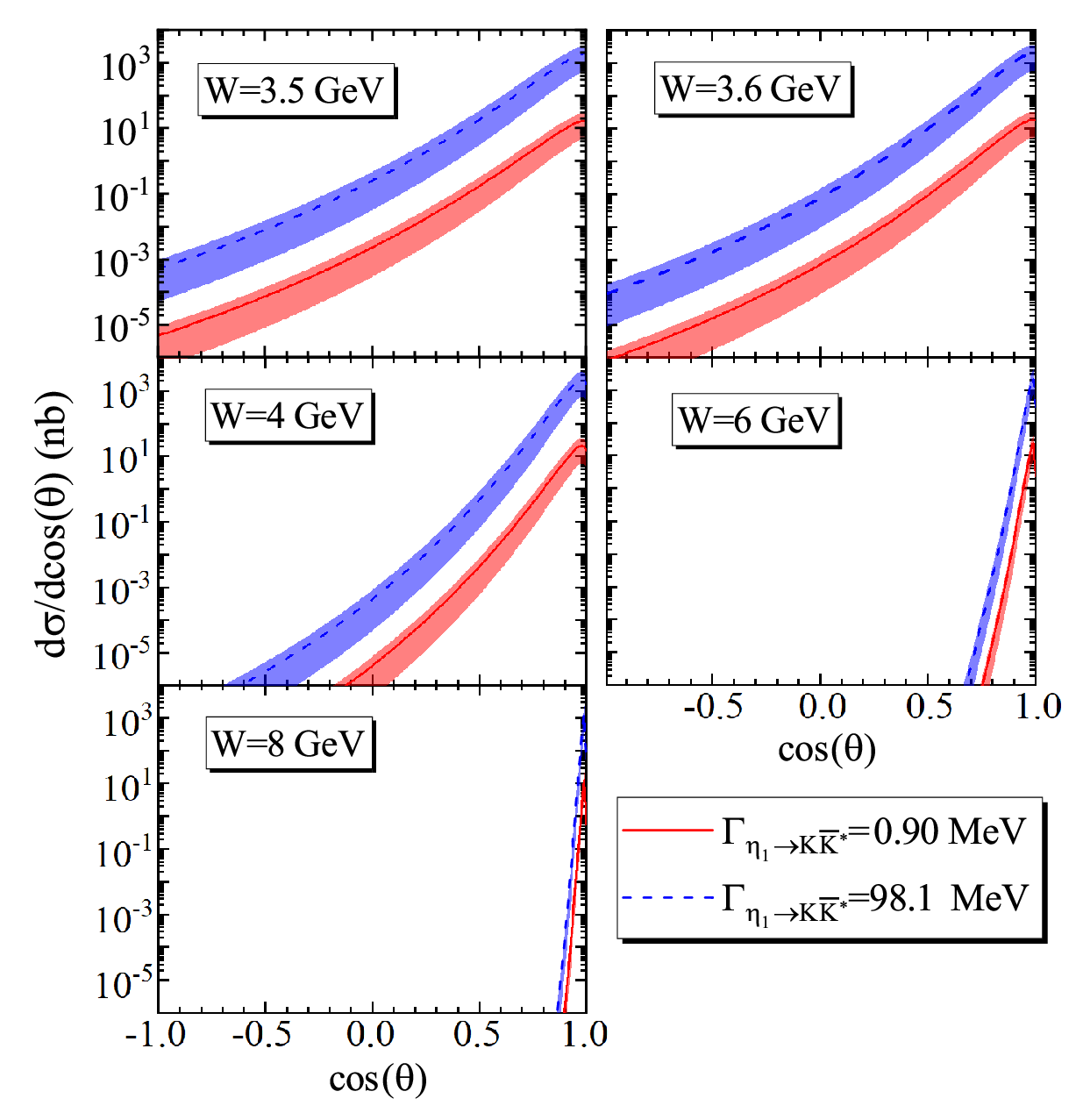}
\caption{(color online).  The differential cross section $d\protect\sigma %
/d\cos \protect\theta $ of the $\protect\eta _{1}(1855)$ production at different c.m. energies
$W=3.5\,{\rm GeV},\,3.6\,{\rm GeV},\, 4,6\,{\rm GeV}$ and $8$ GeV. Here, the notation is the same as that in Fig. \protect\ref
{Fig:total}.}
\label{dcs}
\end{figure}

Since there are no experimental data for the $K^{-}p\rightarrow
\eta_{1}(1855)\Lambda $ reaction, here we give the prediction of the cross section
of the reaction as presented in Figures~\ref{Fig:total}-\ref{dcs}. In
these calculations, the cutoff parameter involved in the form factor is the only free
parameter. In Ref. \cite{Xiang:2020phx}, by fitting the experimental data of
the $\pi ^{-}p\rightarrow K^{\ast }\Sigma ^{\ast }$ process, the $%
\Lambda _{t}=1.67\pm 0.04$ GeV corresponding to the contributions of
Reggeized $t$-channel $K^{(\ast)}$ exchange is obtained. Moreover, in Ref.
\cite{Ozaki:2010wp}, for the Reggeized $t$ channel with the $K$ and $K^{\ast }$
exchanges, the experimental data can be reproduced well by taking cutoff $%
\Lambda _{t}=1.55$ GeV. Also in Ref. \cite{Wang:2018mjz}, the result indicate
that $\Lambda _{t}=1.60$ GeV is a reasonable value for the $%
K^{-}p\rightarrow f_{1}(1420)\Lambda $ process through Reggeized $t$-channel
$K^{\ast }$ exchange. Thus, in this work one intends
to take the value of cutoff for the $t$-channel $K^{\ast}$ exchange to be $%
1.6\pm0.3$ GeV.

In Fig.~\ref{Fig:total}, we present the total cross section of the $%
K^{-}p\rightarrow \eta _{1}(1855)\Lambda $ reaction within the Regge trajectory model by
taking $\Lambda _{t}=1.6\pm 0.3$ GeV. It is found that the
line shape of the total cross sections of the $K^{-}p\rightarrow \eta
_{1}(1855)\Lambda $ process
goes up very rapidly and has a peak around $W=3.4\sim 3.6$ GeV. Taking the partial decay width of the $\eta _{1}$ to $K\bar{K}^{\ast }$ as 0.9 MeV and 98.1 MeV, the peak of the total cross section of the $\eta_1(1855)$
production in the $K$ induced reaction can reach up to about 2.4 $nb$ and 261.5 $nb$ at $%
W=3.4\sim 3.6$ GeV, respectively, which indicates that $W\in( 3.4 \,{\rm GeV},
3.6 \,{\rm GeV})$ is the best energy window for searching for the $\eta _{1}(1855)$ via the kaon induced reaction. Even if we take into account the error of the cutoff value, at the best energy window, the minimum
cross section of the $\eta _{1}(1855)$ production via the $K^{-}p$ reaction are about 0.59 $nb$ and 63.8 $nb$, respectively.
Obviously, using the partial width prediction of $\eta _{1}$ given by the molecular state and hybrid state models \cite{Chen:2022qpd,Dong:2022cuw}, the calculated cross section sizes of $K^{-}p\rightarrow \eta
_{1}(1855)\Lambda$ differ by at least two orders of magnitude.

In Fig.~\ref{dcs}, one presents the prediction of the differential cross section
of the $K^{-}p\rightarrow \eta _{1}(1855)\Lambda $ reaction within the Regge trajectory model by taking
a cutoff $\Lambda _{t}=1.6\pm 0.3$ GeV. It can be seen that the differential
cross sections of the reaction are very sensitive to the $\theta $
angle, which show strong forward-scattering enhancements especially at higher
energies. Thus, the measurement at forward angles is
suggested, which can be applied to check the validity of the Reggeized
treatment.

\begin{figure}[htbp]
\centering
\includegraphics[scale=0.43]{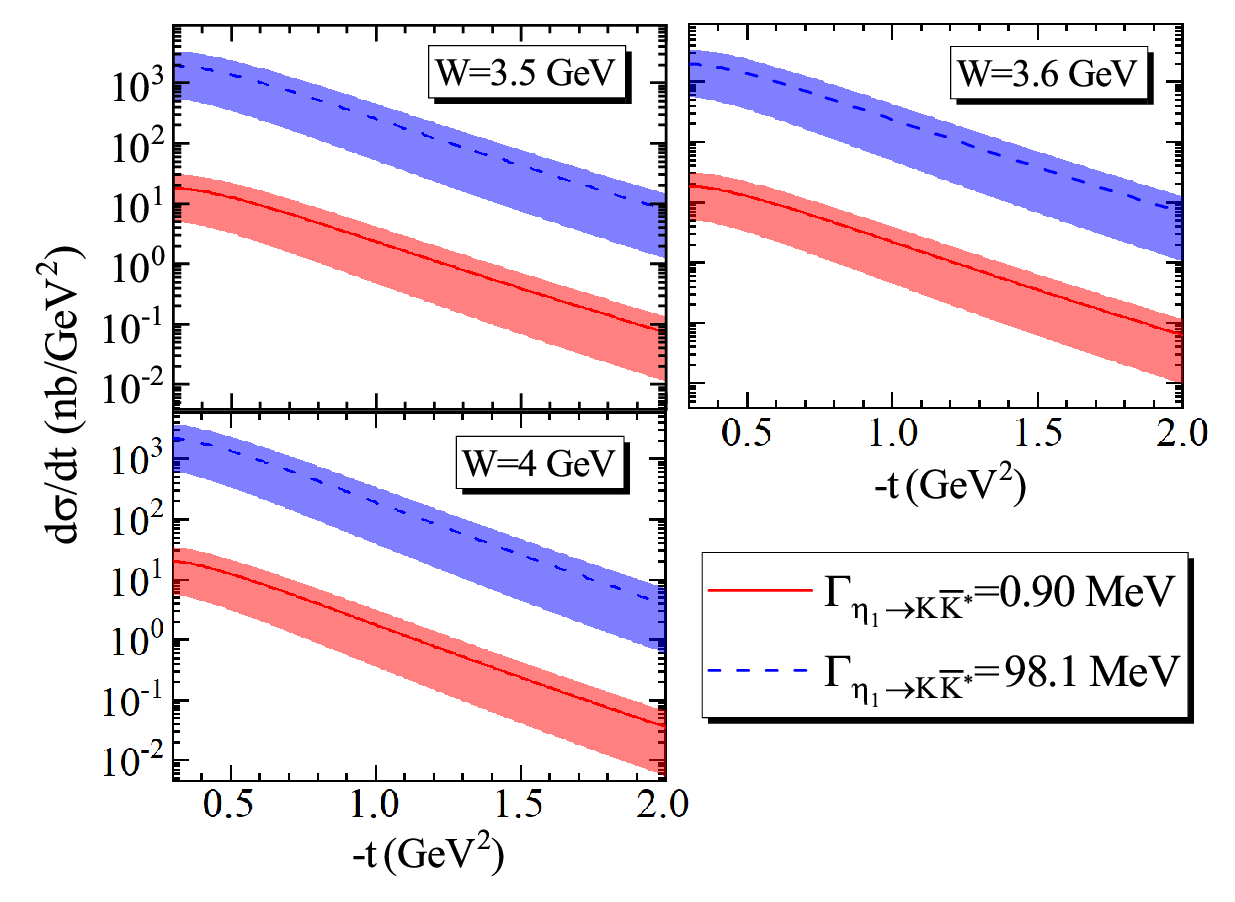}
\caption{(color online).  The $t$-distribution for the $K^{-}p\rightarrow
\protect\eta _{1}(1855)\Lambda $ reaction at different c.m. energies $%
W=3.5{\rm GeV},\, 3.6\,{\rm GeV} $ and $4$ GeV. Here, the notation is the same as that in Fig. \protect\ref%
{Fig:total}. }
\label{tslopeA}
\end{figure}

\begin{figure}[htbp]
\centering
\includegraphics[scale=0.43]{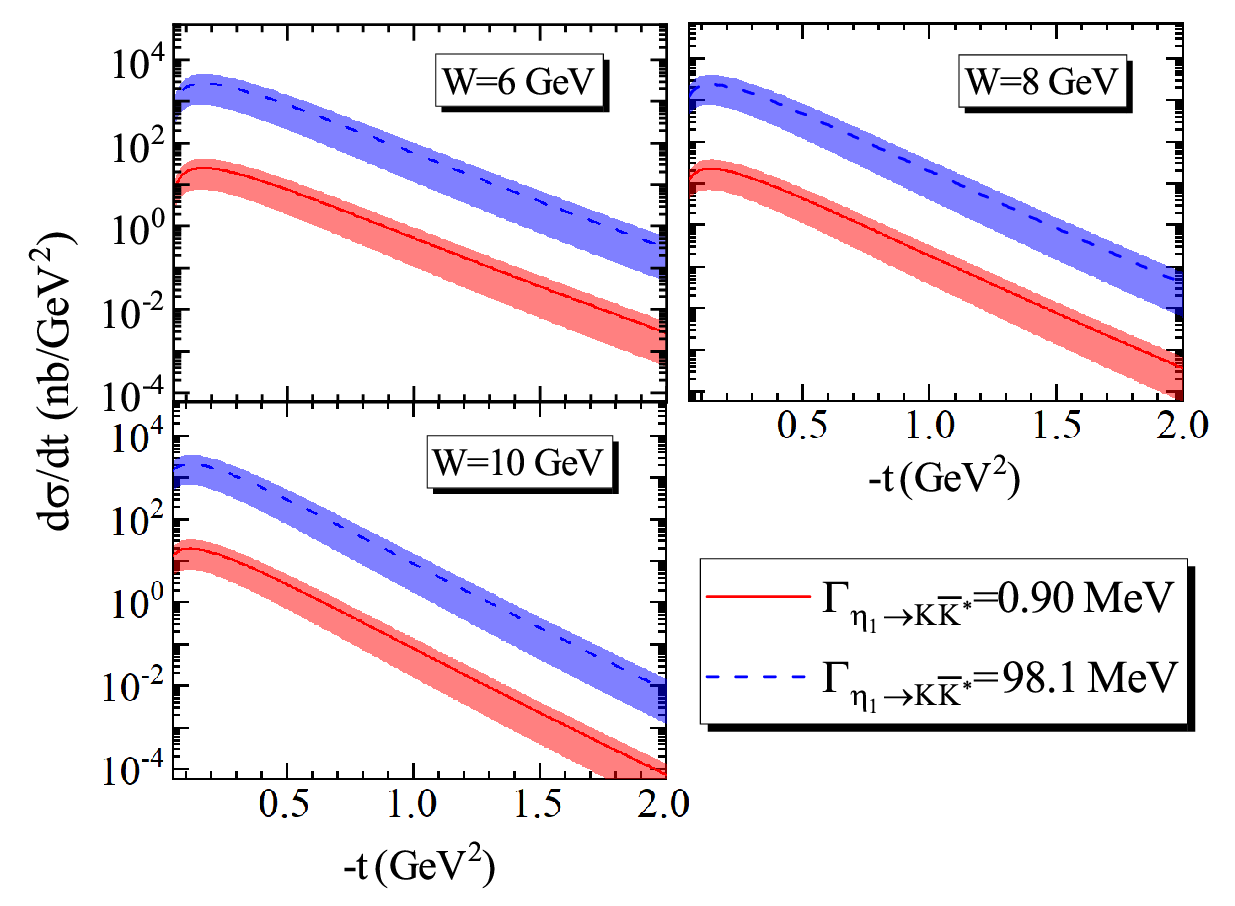}
\caption{(color online).  Same as Fig. \protect\ref{tslopeA} except that the
c.m. energy is expanded to 6 GeV, 8 GeV and 10 GeV.}
\label{tslopeB}
\end{figure}

\begin{figure}[htbp]
\centering
\includegraphics[scale=0.35]{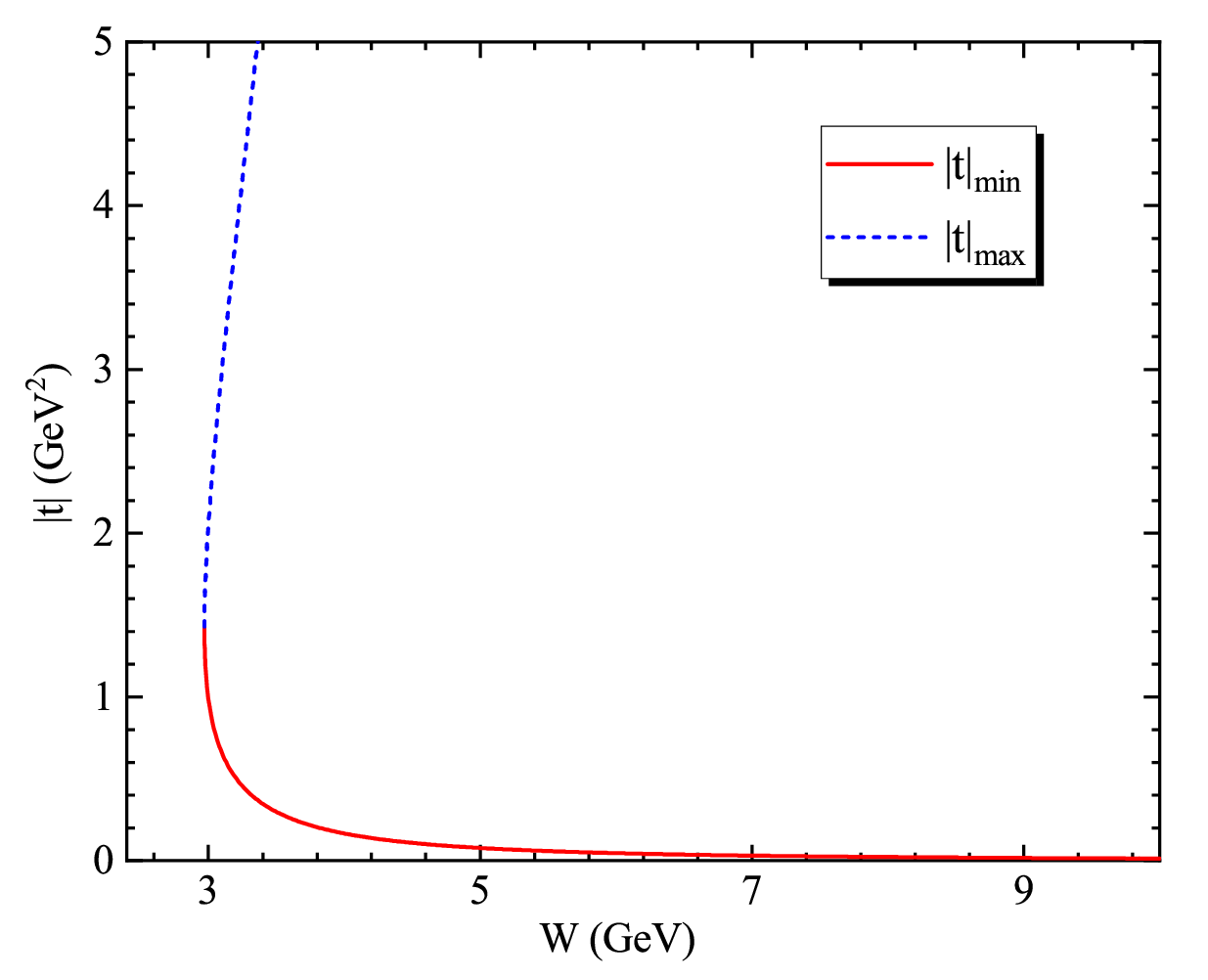}
\caption{(color online). The limiting value of $|t|_{\min }$ as a function of $%
W $. }
\label{tmin}
\end{figure}

In Fig. \ref{tslopeA} and Fig. \ref{tslopeB}, the $t$-distributions for the $%
K^{-}p\rightarrow \eta _{1}(1855)\Lambda $ at low and high
energies are given. As can be seen from both figures, in the
small momentum transfer $t$ region, the shape
of the $t$ distributions of the $K^{-}p\rightarrow \eta _{1}(1855)\Lambda $ appears to be somewhat curved, which may help us
distinguish the production mechanism of the $\eta _{1}(1855)$.
By comparison, one finds that the degree of curvature of the differential cross section at small $t$ increases with the c.m. energy, which is mainly
caused by the limitation of $\left\vert t_{\min }\right\vert $. Figure \ref%
{tmin} shows the change of $\left\vert t_{\min }\right\vert $ value with
energy. When the center of mass energy is 3.5 GeV and 6 GeV, the corresponding $%
\left\vert t_{\min }\right\vert $ values are 0.298  GeV$^{2}$ and 0.046 GeV$^{2}$,
respectively. It means that at low energies, neither experiment nor theory
can give values of $t$ distributions in the region of $t\sim 0$.

\subsection{Dalitz process}

\begin{figure}[h]
\centering
\includegraphics[scale=0.4]{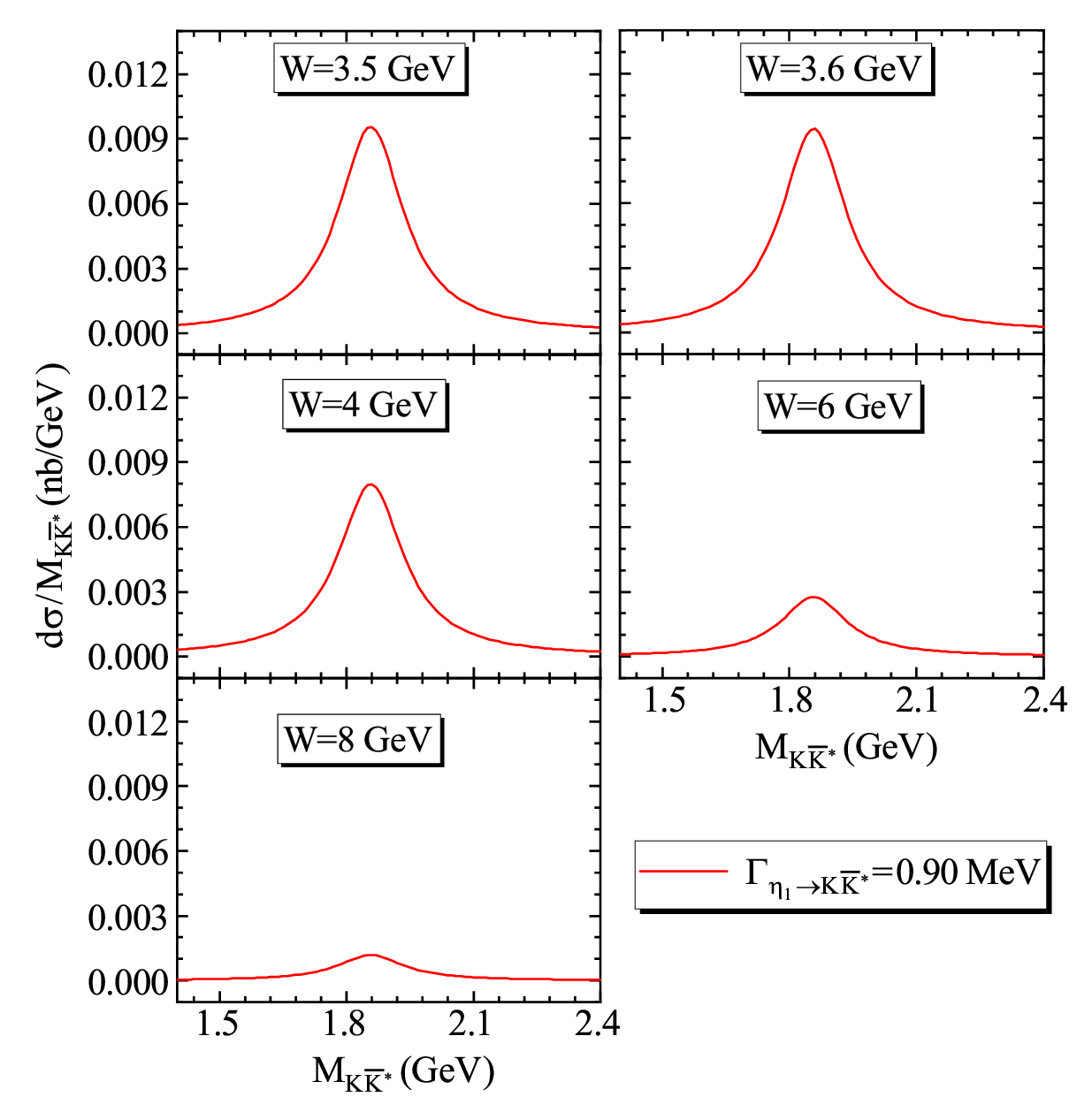}
\caption{The invariant mass distribution $ d\sigma _{K^{-}p\rightarrow \eta _{1} \Lambda \rightarrow K\bar{K}^{\ast } \Lambda }/ dM_{K\bar{K}^{\ast }}$ reaction at different c.m. energies $W=3.5\,{\rm GeV},\,3.6\,{\rm GeV},\, 4\,{\rm GeV},\, 6\,{\rm GeV}$ and $8$ GeV for the partial decay width $\Gamma_{\eta _{1}\rightarrow K\bar{K}^{\ast}}=0.90$ MeV. Here, the cutoff $\Lambda _{t}$ is taken as 1.3 GeV.}
\label{fig:deta}
\end{figure}

\begin{figure}[h]
\centering
\includegraphics[scale=0.4]{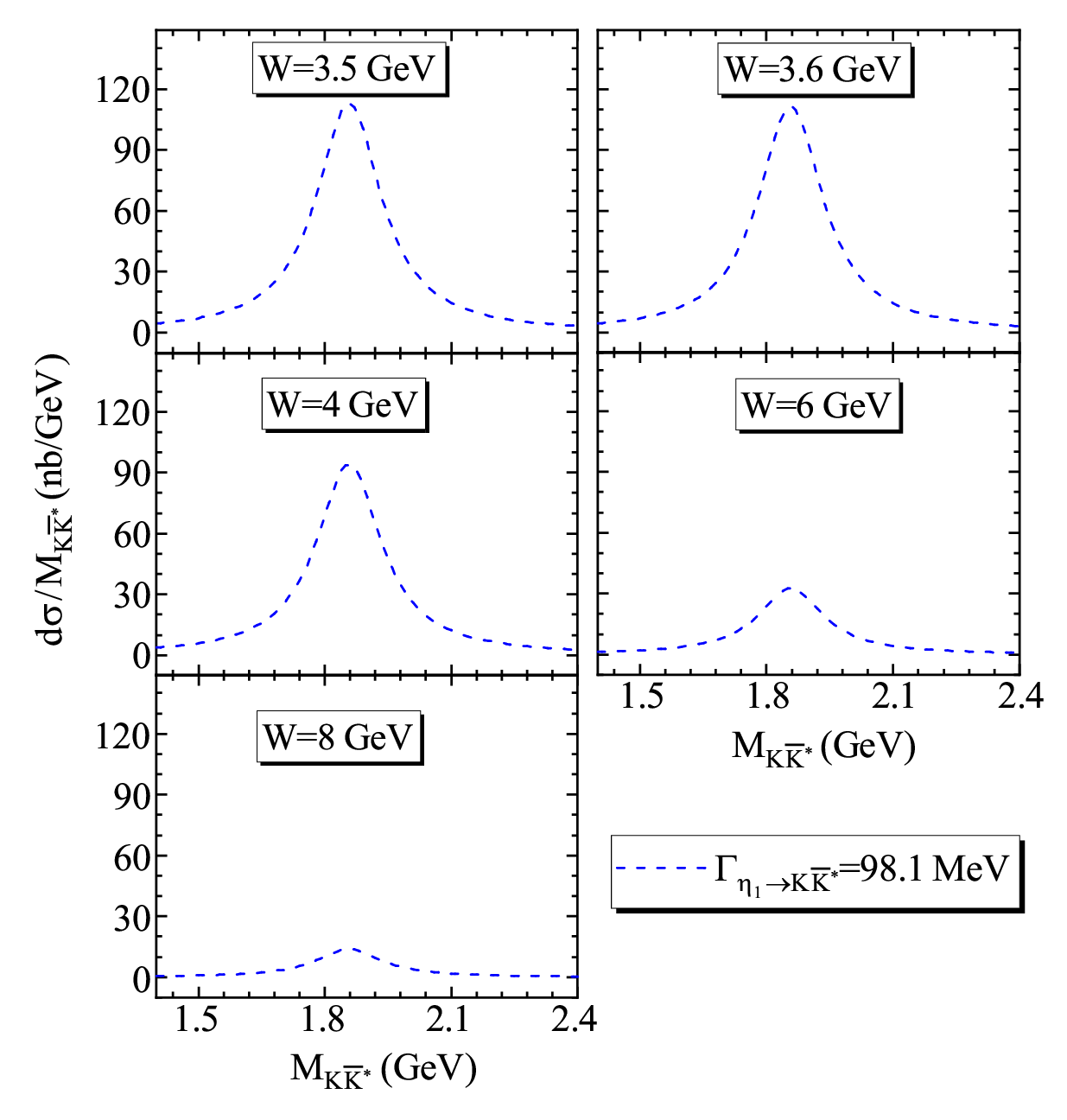}
\caption{Same as Fig. \protect\ref{fig:deta} except that the
$\Gamma_{\eta _{1}\rightarrow K\bar{K}^{\ast}}$ is taken as 98.1 MeV.}
\label{fig:deta2}
\end{figure}

Note that the $\eta _{1}(1855)$ is an intermediate
states usually reconstructed by its decays. And from Table \ref{width}, the
decay widths of the $\eta _{1}(1855)$ to different final
states are different. For example, in the hybrid state hypothesis \cite{Chen:2022qpd}, the width of the $\eta _{1}(1855)$ decays to $K\bar{K}^{\ast }$ is very large, while in the framework of the molecular state model \cite{Dong:2022cuw}, the $\eta _{1}(1855)$ mainly decays to the three body state $K\bar{K}^{\ast }\pi$. Thus, one should consider, for example, the Dalitz process
of $K^{-}p\rightarrow \eta _{1}(1855)\Lambda \rightarrow K \bar{K}^{\ast }
\Lambda $, which may provide a useful reference for future experiments.
Usually, the invariant mass distribution for the Dalitz process $%
K^{-}p\rightarrow \eta _{1}(1855)\Lambda \rightarrow K\bar{K}^{\ast } \Lambda $
can be defined with the two-body process \cite{Kim:2017nxg}%
\begin{equation*}
\frac{d\sigma _{K^{-}p\rightarrow \eta _{1}\Lambda \rightarrow K\bar{K}^{\ast } \Lambda }}{dM_{K\bar{K}^{\ast } }}\approx \frac{2M_{\eta _{1}}M_{K\bar{K}^{\ast }}}{\pi }\frac{\sigma _{K^{-}p\rightarrow \eta _{1}\Lambda
}\Gamma _{\eta _{1}\rightarrow K\bar{K}^{\ast } }}{(M_{K\bar{K}^{\ast }
}^{2}-M_{\eta _{1}}^{2})^{2}+M_{\eta _{1}}^{2}\Gamma
_{\eta _{1}}^{2}}.
\end{equation*}%
where the full width $\Gamma _{\eta _{1} } $ is taken as $188$ MeV. The value of the partial width $\Gamma _{\eta _{1} \rightarrow K\bar{K}^{\ast }} $ is taken as 0.9 MeV and 98.1 MeV, which correspond to the molecular state and hybrid state images, respectively, as listed in Table \ref{width}.
Thus the invariant-mass distribution $ d\sigma _{K^{-}p\rightarrow \eta _{1} \Lambda \rightarrow K\bar{K}^{\ast} \Lambda }/ dM_{K\bar{K}^{\ast}  }$ for $W=3.5-8$ GeV are calculated, as shown in Fig. \ref{fig:deta} and Fig. \ref{fig:deta2}.
It is seen that there exists
an obvious peak at $M_{K\bar{K}^{\ast }}$ near $ 1.86$ GeV. Most notably, the size of the results corresponding to the value of the partial width $\Gamma _{\eta _{1} \rightarrow K\bar{K}^{\ast }} $ varies greatly, which directly affects the difficulty of experimentally measuring the $\eta _{1}(1855)$.

To further investigate the feasibility of experimentally searching for the $\eta _{1}(1855)$ by $\bar{K}p$ scattering, it is necessary to calculate the ratio of $\sigma (K^{-}p\rightarrow \eta _{1}(1855)\Lambda \rightarrow K\bar{K}^{\ast } \Lambda )/\sigma (K^{-}p\rightarrow K\bar{K}^{\ast } \Lambda )$. However, due to the lack of experimental data on $K^{-}p\rightarrow K\bar{K}^{\ast } \Lambda$, and the branching ratio of $K^{\ast }$ decaying to $K\pi$ reaches $100\%$, we extend the process of $K^{-}p\rightarrow K\bar{K}^{\ast } \Lambda$ to the process of $K^{-}p\rightarrow K\bar{K} \pi \Lambda$ for investigation. Under the hybrid model assumption, by taking the $\Gamma _{\eta _{1} \rightarrow K\bar{K}^{\ast }} =98.1 $ MeV in Ref. \cite{Chen:2022qpd}, the minimum cross section of the  $\eta _{1}(1855)$ production via the $K^{-}p$ reaction is about 63.8 $nb$ at $W=3.5$ GeV. With the branching ratio $BR(\eta _{1}\rightarrow K \bar{K} \pi) \approx 52.2\%$, one obtains the total cross section $\sigma _{K^{-}p\rightarrow \eta _{1}\Lambda \rightarrow K  \bar{K} \pi \Lambda } \approx  33.3$ $nb$ at $W=3.5$ GeV. There are several experimental data available for the cross section of the $K^{-}p\rightarrow \Lambda \pi  K\bar{K}$
reaction \cite{data}, which are listed in Table \ref{DATA:I}.

\begin{table}[htbp]
\renewcommand{\arraystretch}{1.50}
\centering
\caption{ The cross section for the $K^-p\to \Lambda \pi  K \bar{K} $
reaction at different beam momentum from several experiments \cite{data}.  }
\begin{tabular}{ccccc}
\toprule[1.0pt]
\toprule[1.0pt]\noalign{\smallskip}
~~~Reaction~~~ &  ~~~$W$ (GeV) ~~~     & ~~~ Cross section ($\mu b$)~~~ \\
 \hline\noalign{\smallskip}
 ~~~ $K^-p\to \Lambda \pi^0 K^0 K^{0} $~~~    & 3.52 &  $39 \pm 14$ \\
 \hline\noalign{\smallskip}
  $K^-p\to \Lambda \pi^0 K^0 K^{0}  $   &3.65 & $39\pm 6$  \\
\bottomrule[1.0pt]
\bottomrule[1.0pt]
\end{tabular}
\label{DATA:I}
\end{table}

In the $K^{-}p\rightarrow \Lambda \pi  K \bar{K}$ process, we estimate that the total cross section  is   39 $\mu b$ near the c.m. energy $W=3.5$ GeV according to the average cross section of these energies. Therefore, for the hybrid model, one get
the ratio at $W=3.5$ GeV  as
 \begin{equation}
\frac{\sigma (K^{-}p\rightarrow \eta _{1}\Lambda \rightarrow K\bar{K}\pi \Lambda  )}{\sigma (K^{-}p\rightarrow K  \bar{K}\pi \Lambda  )}\simeq 0.085\%.
\end{equation}
When checking the present experimental status, we notice that the J-PARC experiment can provide kaon meson beams \cite{Nagae:2008zz,jparc}, which is an ideal facility for probing the $\eta _{1}(1855)$ production by the $\bar{K}p$ reaction. Usually, the rate of $K$ to $\pi$, both produced as secondary beams in J-PARC, is less than $1/2$ at $\eta _{1}(1855)$ production energies.
According to the experimental status of J-PARC \cite{Aoki:2021cqa}, it can be expected to detect about $47,000$ events for the $K  \bar{K}\pi \Lambda$ production per 100 day, in which about dozens of events are related to the $\eta _{1}(1855)$. This means that it is feasible for us to search for the $\eta _{1}(1855)$ by $\bar{K}p$ scattering experiments, of course, the basis of this estimate is based on the assumption that $\eta _{1}$ is a hybrid state \cite{Chen:2022qpd}. For the molecular state model \cite{Dong:2022cuw}, we calculate $\sigma _{K^{-}p\rightarrow \eta _{1}\Lambda \rightarrow K  \bar{K} \pi \Lambda } \approx  0.0028$ $nb$ at $W=3.5$ GeV, which is four orders of magnitude smaller than the case of the hybrid state assumption, which would make experimental detection of $\eta _{1}$ very difficult.

Moreover, it can be seen from Table  \ref{width} that $\eta _{1}$ mainly decays to $K\bar{K}^{\ast }\pi$ under the molecular state image. The experimental data for the $K^{-}p\rightarrow K  \bar{K}^*\pi \Lambda$ reaction are listed in Table \ref{DATA:II}, and we can estimate that the average cross section of $K^{-}p\rightarrow K  \bar{K}^*\pi \Lambda$ at $W=3.5$ is about 13.8 $\mu b$. Taking a similar analysis method, considering that the minimum cross section of the $\eta _{1}(1855)$ production is above 0.59 $nb$ near the $W=3.5$ GeV and the branching ratio $BR(\eta _{1} \rightarrow K  \bar{K}^*\pi)=55.9\%$,  one get the
the cross section ratio at $W=3.5$ GeV  as
  \begin{equation}
\frac{\sigma (K^{-}p\rightarrow \eta _{1} \Lambda \rightarrow K  \bar{K}^*\pi \Lambda  )}{\sigma (K^{-}p\rightarrow K  \bar{K}^*\pi \Lambda  )}\simeq 0.0024\%.
\end{equation}
Obviously, it is a very small value, which means that it is almost impossible for us to search for the $\eta _{1}(1855)$ in J-PARC experiments through the $K^{-}p\rightarrow K  \bar{K}^*\pi \Lambda$ process.
\begin{table}[htbp]
\renewcommand{\arraystretch}{1.50}
\centering
\caption{ The cross section for the $K^-p\to \Lambda \pi  K \bar{K}^{* } $
reaction at different beam momentum from several experiments \cite{data}.  }
\begin{tabular}{ccccc}
\toprule[1.0pt]
\toprule[1.0pt]\noalign{\smallskip}
~~~Reaction~~~ &  ~~~$W$ (GeV) ~~~     & ~~~ Cross section ($\mu b$)~~~ \\
 \hline\noalign{\smallskip}
 ~~~ $K^-p\to \Lambda \pi^+ K^0 K^{*-} $~~~    & 3.52 &  $13 \pm 6$ \\
 \hline\noalign{\smallskip}
  $K^-p\to \Lambda \pi^0 K^0 K^{*0}  $   &3.52 & $12\pm 9$  \\
 \hline\noalign{\smallskip}
    $K^-p\to \Lambda \pi^-  K^+ K^{*0} $  & 3.52  & $16.5\pm 7.5$  \\
\noalign{\smallskip}
\bottomrule[1.0pt]
\bottomrule[1.0pt]
\end{tabular}
\label{DATA:II}
\end{table}

\section{Summary}\label{sec4}

Recently, the BESIII Collaboration reported the observation of  the $\eta_1(1855)$ in the $J/\psi\to \gamma\eta\eta^\prime$ decay \cite{BESIII:2022riz}. Since the $\eta_1(1855)$ is an isoscalar resonance with exotic $J^{PC}=1^{-+}$ quantum numbers,
different explanations of exotic states like hybrid \cite{Chen:2022qpd,Qiu:2022ktc,Shastry:2022mhk}, the $K\bar{K}_1(1400)$ molecular state \cite{Dong:2022cuw,Yang:2022lwq}, and tetraquark state \cite{Wan:2022xkx} were proposed, which shows that this observation attracted theorist's attention in the past months.
Besides decoding the nature of the observed $\eta_1(1855)$, establishing the $\eta_1(1855)$ in experiment is also crucial.
How to confirm the observation of the $\eta_1(1855)$ by other experiments becomes a central issue.

In this work, we propose that the $\bar{K}p$ reaction can be one way to further explore the $\eta_1(1855)$, which is stimulated by the study of its strong decays \cite{Chen:2022qpd,Dong:2022cuw}. Our calculation provides the information of the total and differential
cross sections of the $K^{-}p\to\eta _{1}(1855)\Lambda$ reaction. Using the predicted partial width of $\eta_1$ in the hybrid model \cite{Chen:2022qpd}, our results show that searching for the $\eta_1(1855)$ through the $K^{-}p\rightarrow \eta _{1}\Lambda \rightarrow K\bar{K}^{\ast } \Lambda \rightarrow K\bar{K}\pi \Lambda$ process is a feasible way. However, according to the predicted width of the molecular state model \cite{Dong:2022cuw}, calculations show that it is extremely difficult to observe the $\eta_1(1855)$ through the $\bar{K}p$ scattering process. Note that the current calculation of the cross section of the $\eta_1(1855)$
through the $\bar{K}p$ process still relies on the predicted value of the decay partial width given by the theoretical models \cite{Chen:2022qpd,Dong:2022cuw}. Thus more in-depth theoretical model predictions are needed, which is of great significance to reduce the uncertainty and provide reliable reference for future experiments.

In summary, we suggest these available experiments at OKA@U-70~\cite{Obraztsov:2016lhp} and SPS@CERN~\cite%
{Velghe:2016jjw}, and J-PARC~\cite{Nagae:2008zz,jparc} to explore the newly reported $\eta_1(1855)$, which are of interest to clarify the internal structural properties of the $\eta_1(1855)$.

\section{Acknowledgments}

This project is supported by the National Natural Science Foundation of China under Grant Nos. 12065014 and 12047501,
and by the West Light Foundation of The Chinese Academy of Sciences, Grant No. 21JR7RA201. X.L. is also supported by the China National Funds for Distinguished Young Scientists under Grant No. 11825503, National Key Research and Development Program of China under Contract No. 2020YFA0406400, and the 111 Project under Grant No. B20063, the Fundamental Research Funds for the Central Universities, and Science and Technology Department of Qinghai Province Project No. 2020-ZJ-728.

\end{document}